
\documentclass[twocolumn]{aastex631} 
\usepackage{graphicx}
\usepackage{float}
\usepackage{textcomp}
\usepackage{epstopdf}
\usepackage{soul}
\usepackage{xcolor}
\usepackage{hyperref}
\usepackage{lineno}
\usepackage{float}

\submitjournal{The Astrophysical Journal}

\shorttitle{Exotrojans}
\shortauthors{Wood and Cummins}

\newcommand{\textedit}{}

\newcolumntype{Y}{>{\centering\arraybackslash}X}

\begin{document}
\title{Three Body Mean Motion Resonance Chains as a Delivery Mechanism for White Dwarf Pollution}

\correspondingauthor{Jeremy Wood}

\author[0000-0003-1584-302X]{Jeremy Wood}
\author{Natalie Cummins}
\affiliation{American Military University, 111 W. Congress Street, Charles Town, WV 25414}

\begin{abstract}
In this work, we used numerical integration of the 4-body problem to study \textedit{3-body resonance chains} (two planets and an asteroid in the innermost orbit) as a possible mechanism for white dwarf pollution. Two \textedit{3-body resonance chains} were selected for study: the 6:3:2 and the 4:2:1. Asteroids in both a dynamically colder initial orbit in the 6:3:2 resonance and hotter initial orbits in both resonances were studied. An asteroid had up to a 1.08$\%$ chance of being delivered to the stellar Roche Limit of the white dwarf. This probability was strongly linearly correlated with the mass of the inner planet but was not correlated with the mass of the outer planet for both colder and hotter orbits. \textedit{Average dynamical lifetimes ranged from 23 kyr to 1137 kyr for the dynamically colder orbit and from 12.9 kyr to 89.2 kyr and 10.8 kyr to 793.4 kyr for the dynamically hotter orbits in the 6:3:2 and 4:2:1 resonances, respectively. Average dynamical lifetime was exponentially anticorrelated with the outer planet mass and usually with the inner planet mass except in one case.} The \textedit{hotter} 4:2:1 resonance delivered 1.1 times more asteroids to the stellar Roche Limit than the \textedit{hotter} 6:3:2 resonance. The \textedit{hotter} 6:3:2 resonance delivered 1.2 times more asteroids to the stellar Roche Limit than the colder 6:3:2 resonance. A typical accretion rate for a white dwarf star of 10$^8$ grams s$^{-1}$ could be explained by the accretion of an equivalent mass of one of our simulated asteroids every 13.8 Myr.
\end{abstract}



\section{Introduction}
White dwarf stars are the leftover cores of medium-sized stars that have shed their outer layers during the asymptotic giant branch phase after having left the main sequence and evolving over time. Their sizes are comparable to the size of Earth, and their masses are below 1.4 solar masses \citep[e.g.][]{1939isss.book.....C,1996imsa.book.....O}.\par
The atmosphere of a white dwarf star should be dominated by either hydrogen and$\backslash$or helium \citep{1979ApJ...231..826F,1986ApJS...61..197P,2014MNRAS.439.3371W}. However, observations of white dwarf spectra show that over 1,000 white dwarf stars are polluted with metals and volatiles such as beryllium, calcium, iron, magnesium, potassium, lithium, water ice, and much more \citep{1992ApJS...82..505D,1993ApJS...84...73D,2003ApJ...596..477Z,2004ApJ...607..426K,2006ApJS..167...40E,2007ApJ...663.1291D,2010ApJ...722..725Z,2013ApJS..204....5K,2014AREPS..42...45J,2015MNRAS.446.4078K,2016MNRAS.455.3413K,2017MNRAS.467.4970H,2018MNRAS.479.3814H,2019MNRAS.487..133W,2020MNRAS.499..171H,2021ApJ...914...61K,2021MNRAS.508.5671L,2022MNRAS.517.4557E,2023MNRAS.519.2646B,2023MNRAS.519.2663B}. The most likely origin of this pollution is planetesimals that have become tidally disrupted due to the crossing of their stellar Roche Limit \citep{2010ApJ...722..725Z,2014AREPS..42...45J,2014MNRAS.439.3371W,2016NewAR..71....9F,2020MNRAS.499.1854M,2024MNRAS.529.2910N,2024MNRAS.532.3866R}.\par
Since the cooling ages of polluted white dwarfs can far exceed the sinking times of the polluting debris by orders of magnitude, the pollution must be recent or have a replenishing source \citep[][]{2006A&A...453.1051K,2009A&A...498..517K,2018ApJ...863..184B,2022MNRAS.517.4557E}. \par
Different mechanisms have been proposed to explain how bodies could be brought inward to pollute the star. These include wide stellar binaries \citep[][]{2015MNRAS.454...53B}, eccentric planets \citep[][]{2014MNRAS.439.2442F}, exomoons \citep[][]{2016MNRAS.457..217P,2017MNRAS.464.2557P,2022ApJ...936...30T}, instability in two-planet systems \citep[][]{2013MNRAS.431.1686V}, secular resonances \citep[][]{2017ApJ...834..116P,2018MNRAS.480...57S,2021MNRAS.504.3375S}, and 2-body mean motion resonances \citep[][]{2012ApJ...747..148D,2016MNRAS.463.4108A,2021MNRAS.504.3375S}. \par
In this work, we investigate the efficiency of \textedit{3-body resonance chains} that contain two planets in a strong 2-body mean motion resonance with each other and an asteroid in delivering the asteroid to its stellar Roche Limit in a white dwarf star system where it could be torn apart by tidal forces and eventually pollute the atmosphere of the white dwarf. \textedit{3-body resonance chains} between two planets and a small body abound in our Solar system \citep[][]{2014Icar..231..273G}, but perhaps the classic case of a \textedit{3-body resonance chain} in our own Solar system is the 4:2:1 resonance among Jupiter's moons Europa, Ganymede, and Io which is also known as a Laplace resonance \citep[e.g.][]{1999ssd..book.....M,2024CeMDA.136...19L}.\par
But no \textedit{3-body resonance chains} with planets in a strong 2-body mean motion resonance (having orbital periods in a ratio of small integers) exist in our own Solar system. Though there are planets in our own Solar system that are nearly in a strong 2-body mean motion resonance with another planet (e.g. the near 5:2 mean motion resonance between Saturn and Jupiter; the near 2:1 mean motion resonance between Neptune and Uranus). However, N-body mean motion \textedit{resonance chains} among planets can be found in exoplanet systems.\par 
Examples include \textedit{2-planet mean motion resonances} \citep[][]{2004A&A...415..391M,2012ApJ...754...50R,MMR3,2016ApJ...818...35W,2019AJ....158..133H}, a \textedit{3-planet resonance chain} \citep[][]{2010ApJ...719..890R,2018AJ....155..106M}, a \textedit{5-planet resonance chain} \citep[][]{2021A&A...649A..26L}, and a \textedit{6-planet resonance chain} \citep[][]{2023Natur.623..932L,2024ApJ...968L..12L}. Thus, it is possible for undiscovered asteroids in exoplanet systems to exist in N-body \textedit{resonance chains} with planets that are in mean motion resonances with each other. Furthermore, since white dwarf star systems are capable of supporting ancient planets that have survived the star's two red giant phases \citep[][]{2010ApJ...722..725Z}, it is also possible for white dwarf star systems to contain planets in mean motion resonances with each other. During a star's asymptotic giant branch phase, mass loss of the star and possible mass gain of a planet can increase the width of mean motion resonances, trapping asteroids that were previously outside the resonance \citep[][]{2012ApJ...747..148D}, creating the possibility of a strong \textedit{3-body resonance chain}.\par
The motivation for our work comes from \citet{1982AJ.....87..577W} who observed that asteroids in a 2-body mean motion resonance with Jupiter can exhibit large spikes in eccentricity. Our investigation will determine the effect of the mass of each planet starting in a 3-body mean motion resonance \textedit{chain} with an asteroid on the efficiency of delivering the asteroid to its stellar Roche Limit.\par
This paper is partitioned as follows: in section 1 we introduce the topic. In section 2 we explain our methodology. In section 3 we present our results, and finally in section 4 we draw conclusions.

\section{Method}
We used the technique of numerical integration in the 4-body problem. In effect, our system consisted of a white dwarf star, a massless test particle to represent an asteroid, and two planets. For this project, we desired the two planets and the asteroid to initially be in a 3-body mean motion resonance. We selected two resonances for study: the 4:2:1 resonance (the Laplace resonance) and the 6:3:2 resonance.\par
We define a simulation as consisting of two planets, a white dwarf star, and 10,000 massless non-interacting test particles. Using massless test particles to simulate small bodies has become an acceptable simplification \citep[e.g.][]{1961ZA.....51..201P,2009MNRAS.398.1715L,2010MNRAS.405...49H,2016A&C....16...26W,2014ApJ...796...23G,2018AJ....155....2W,2022ApJ...929..157W,2023MNRAS.519..812W}. \par 
The mass of our star was set to 0.7 solar mass because this is the average mass of a polluted white dwarf \citep[][]{2014A&A...566A..34K}. This corresponds to a main sequence mass of about 3 solar masses \citep[][]{2008MNRAS.387.1693C,2020MNRAS.499.1854M}. For every simulation, the density of the inner planet was set to that of Jupiter's (1.326 gram cm$^{-3}$ \citep[][]{NASAHorizonsSystem}), and the radius of the outer planet was set to a constant value of approximately 58,300 km which is comparable to the mean radius of Saturn \citep[][]{NASAHorizonsSystem}. The radius of the inner planet was determined from its density and mass. The different masses used for each planet will be discussed in subsequent sections.\par
For each simulation, the initial semi-major axis of the inner planet was set to 10 au so it would be sure to have survived the two red giant phases of the star \citep[][]{2002ApJ...572..556D,2007ApJ...661.1192V,2009ApJ...705L..81V,2012ApJ...761..121M,2014ApJ...794....3V}. Then the initial semi-major axis of the outer planet was always initially set to 15.87 au so that it would always be in a 2:1 mean motion resonance with the inner planet. All other planetary orbital parameters were initially set to zero.\par
The 10,000 test particles were placed in the same orbit \textedit{with orbital parameters discussed in subsequent sections. We studied two dynamically hotter orbits and one dynamically colder orbit.}  \par
Simulations were integrated forward in time. The step time was set to 1 day, and the output time (the time interval at which orbital elements were inspected) was set to 1,000 years.\par
Test particles were removed from the system upon colliding with a planet, crossing the stellar Roche Limit, or achieving an astrocentric distance greater than 100 astronomical units \citep[][]{2014MNRAS.439.2442F}. If a test particle achieved the latter then it was considered to be ejected from the system.\par
The equation for the Roche Limit, $R_{roche}$, for the ideal case of a body held together only due to gravity in a circular orbit and tidally locked is
\begin{equation}
R_{roche} = 2.44R_{*}\bigg{(}\frac{\rho_*}{\rho_a}\bigg{)}^{1/3}
\label{roche_eqn}
\end{equation}

\noindent \citep[][]{Roche_1849,2013ApJ...773L..15R}. Here, $R_{*}$ and $\rho_*$ are the radius and density of the star, respectively. $\rho_a$ is the density of the asteroid. We selected a simulated asteroid density of 1.3 grams cm$^{-3}$. This is slightly larger than the density of the rubble-pile asteroids Bennu and Ryugu (1.19 grams cm$^{-3}$) and less than the density of the rubble-pile asteroid Itokawa (1.9 grams cm$^{-3}$)\citep[][]{2006epsc.conf..493A,2022Icar..38014969S}. \par
For the size of our simulated asteroids we chose a radius of 200 km which is used by the integrator in determining if a collision has occurred. This is large enough so that a debris disk could be created after the asteroid is ripped apart by stellar tidal forces \citep[][]{2003ApJ...584L..91J,2010ApJ...722.1078M,2014AREPS..42...45J,2017MNRAS.468..154B,2017MNRAS.468.1575B,2021MNRAS.508.5671L}. Table \ref{250moons} shows some small bodies of comparable size in our own Solar system.\par

\begin{deluxetable}{ccc}
\tablewidth{\columnwidth}
\tablecaption{\\Bodies with Radii near 200 km\label{250moons}}
\tablehead{\colhead{Body} & \colhead{Radius} & \colhead{Type} \\ 
\colhead{} & \colhead{(km)} & \colhead{} } 

\startdata
    Enceladus&252&Saturn moon\\
    Mimas&198.2&Saturn moon\\
    Miranda&235.8&Uranus moon\\
    10 Hygiea&203.56&Asteroid\\
\enddata

\tablecomments{Small Solar system bodies similar in size to our simulated asteroids which have a radius of 200 km \citep[][]{NASAHorizonsSystem}.}

\end{deluxetable}

To determine the stellar Roche Limit for our simulated asteroids we needed to find the size of our white dwarf. We used the mass-radius relation for white dwarfs \citep[][]{Princeton UniversityWeb} to determine $R_*$ and obtained a value of $R_*=$ 0.0112 solar radius (or about 1.2 Earth radii). We then calculated the density of the white dwarf using its mass and radius and obtained a value of $7.02 \times10^5$ grams cm$^{-3}$. We then used equation \ref{roche_eqn} to find the stellar Roche Limit of our simulated asteroids, and the result was $R_{roche} = $ 2.23 solar radii. \textedit{We used the average dynamical lifetime of test particles in a simulation to measure the stability of the orbit of the test particles in each simulation.}

\subsection{Dynamically Hotter Orbits}
The initial semi-major axis of each \textedit{dynamically hotter} orbit of the test particles was set so that it would be in a 2:1 or 3:2 inner mean motion resonance with the inner planet, depending on the resonance under investigation. From Kepler's 3rd Law, the inner 2:1 mean motion resonance was found to be at 6.30 au, and the inner 3:2 resonance was found to be at 7.63 au. Each orbit of the test particles was placed in the same plane as that of the planets. Test particles were evenly distributed in true anomaly from 0 to 360 degrees.\par
\textedit{To determine the eccentricity of these two orbits, we started by setting} the initial minimum orbit intersection distance, or MOID \citep[e.g.][]{2020A&A...633A..22H}, between the orbit of the inner planet and that of the test particles equal to the Hill Radius \citep[][]{Hill_1878,1999ssd..book.....M,2019dsss.book.....W} of an inner planet with a mass equal to 10 Earth masses (or $\approx 0.03$ Jupiter mass, $M_J$). A mass of 10 Earth masses was selected for the inner planet mass for this purpose because it is the maximum mass of a super-Earth \citep[][]{2019ConPh..60...63I} and is massive enough to perturb asteroids onto star-grazing orbits \citep[][]{2023MNRAS.519.6257V}. The Hill radius, $R_H$, is given by the equation:

\begin{equation}
    R_H = a_p\Bigg{(}\frac{m_p}{3M_*}\bigg{)}^{1/3}
\label{hill_radius_eqn}
\end{equation}

\citep[e.g.][]{1999ssd..book.....M}.
Where $a_p$ and $m_p$ are the semi-major axis of the orbit and mass of the planet, respectively, and $M_*$ is the mass of the star. Using equation \ref{hill_radius_eqn}, the required MOID was found to be about 0.243 au.\par
We then set the MOID \textedit{of each dynamically hotter orbit} equal to the difference between the initial apoastron distance of the orbit of the test particles and the initial periastron distance of the orbit of the inner planet. This method works because the orbits are in the same plane. Thus, the \textedit{initial} MOID was given by the equation:

\begin{equation}
    MOID = a_p(1-e_p) - a(1+e)
\label{specificmoid}
\end{equation}

Where $e_p$ is the initial eccentricity of orbit of the inner planet, which is always zero, and $a$ and $e$ are the initial semi-major axis and eccentricity of the \textedit{dynamically hotter} orbit of the test particles, respectively. We then solved equation \ref{specificmoid} for $e$ and obtained a value of 0.279 for the 6:3:2 resonance and a value of 0.549 for the 4:2:1 resonance. \textedit{All other orbital parameters of the dynamically hotter orbits were initially set to zero.} \textedit{The initial periastron distance for the 6:3:2 resonance was 5.50 au and that for the 4:2:1 resonance 2.84 au.}\par
To determine the effect of the inner planet mass on the number of test particles delivered to the stellar Roche Limit in a simulation, we ran simulations in which we varied the mass of the inner planet among these four values: 10 Earth masses (henceforth $\approx0.03 M_J$), 0.3$M_J$, 0.45$M_J$, and 0.6$M_J$ while holding the mass of the outer planet constant - first at $M_J$, and then at 4$M_J$.\par
\textedit{In our test runs, we measured the shift in the location of the resonance between the inner planet and test particles in dynamically hotter orbits. It is well known that the semi-major axis and eccentricity of two bodies in a mean motion resonance can oscillate about a constant value over time \citep[e.g.][]{1982AJ.....87..577W,2019dsss.book.....W}. This would have the effect of changing the periastron distance of the inner planet, allowing for the possibility of more severe close encounters between test particles and the inner planet.}\par
\textedit{We found that the eccentricity reached a maximum value of 0.175 for the most extreme case of the most massive outer planet paired with the least massive inner planet. For the least extreme case with the most massive inner planet paired with the least massive outer planet, the eccentricity of the inner planet reached a maximum of 0.106. In both cases, the orbit of the inner planet crossed both our dynamically hotter orbits. The resonance location due to the inner planet, either 2:1 or 3:2, varied by 1.5$\%$ and 4.33$\%$ for the least and most extreme cases for each resonance, respectively. The semi-major axis and eccentricity of the inner planet oscillated with a period of 1,691 years in the least extreme case and 764 years in the most extreme case. The resonance location due to the outer planet, either 4:1 or 3:1, varied by 0.48$\%$ for the least extreme case and 0.025$\%$ for the most extreme case. The inner planet also crossed the orbit of the dynamically colder orbit in both cases.}\par
To determine the effect of the outer planet mass on the number of test particles delivered to the stellar Roche Limit in a simulation, we ran simulations in which we varied the mass of the outer planet among these four values: $M_J$, 2$M_J$, 3$M_J$, and 4$M_J$ while holding the inner planet mass constant - first at $\approx0.03M_J$ and then at 0.6$M_J$. This yielded a total of 12 \textedit{unique} simulations per resonance or 24 unique simulations overall.\par 
Simulations were integrated using the IAS15 integrator selected from the \textsc{rebound} suite of N-body integrators \citep{2012A&A...537A.128R,2015MNRAS.446.1424R}. This non-symplectic high-precision integrator is well able to handle close encounters between a planet and a small body by altering the step time during close encounters (when the distance between bodies is $<$ 3$R_H$) to achieve machine precision. This integrator obeys Brouwer's Law \citep[][]{1937AJ.....46..149B} for error accumulation as discussed in \citet{2023MNRAS.519..812W}.\par
To determine the integration time, a series of pre-simulations were integrated, consisting of up to 1,000 test particles. Based on the results of these pre-simulations, it was decided that an integration time of 10 Myr would be more than long enough to obtain quality data for these dynamically hotter orbits.\par
\textedit{To find an estimate of the largest errors due to the integrator}, we set the inner and outer planet mass equal to $\approx0.03M_J$ and 4$M_J$, respectively. Then we integrated our system forward in time from a time of zero to a time of 5 Myr with no test particles and then integrated it backwards in time to the initial time of zero for a total integration time of 10 Myr. At both times of zero (start and finish time) we recorded the total angular momentum and energy of the system along with the mean longitude of each planet.\par
We measured the fractional error in the angular momentum of the system, $\Delta L_{frac}$, using: 

\begin{equation}
\Delta L_{frac}=\frac{|L_f - L_i|}{L_i}
\label{frac_L_change}
\end{equation}

\noindent where $L_i$ and $L_f$ are the magnitudes of the initial and final angular momentum of the system, respectively. We also measured the fractional error in energy of the system in the same way. We then calculated the difference between the initial and final mean longitudes for each planet to \textedit{estimate} errors in planetary positions \citep[e.g.][]{2022ApJ...929..157W}. \par
To determine the effect of density we performed integrations using simulated asteroid densities of 0.6, 2.0, and 3.0 grams cm$^{-3}$ \citep[][]{2006LPI....37.2214B}. We used the parameters for our dynamically hotter orbit starting the test particles in a 6:3:2 resonance with our two planets as previously discussed. The masses of the two planets were $\approx 0.03M_J$ for the inner planet and $M_J$ for the outer planet. We compared the results from these simulations to the analogous simulation that used our selected simulated asteroid density of 1.3 grams cm$^{-3}$. We found that changing the density had almost no effect on the overall results.

\subsection{Dynamically Colder Orbit}
We also integrated simulations of \textedit{10,000} test particles initially in the 6:3:2 resonance using the same integrator, \textedit{initial conditions, and orbital parameters} used for the dynamically hotter orbit \textedit{in the 6:3:2 resonance} but \textedit{with one crucial difference.} The initial eccentricity of the orbit of the test particles was set to a much lower value of 0.05, \textedit{making it dynamically colder. We solved equation \ref{specificmoid} for the initial MOID with the inner planet and obtained a value of 1.99 au. The initial periastron distance of this colder orbit was 7.25 au.} \textedit{Furthermore,} the total integration time was increased to 100 Myr \citep[][]{2014MNRAS.439.2442F,2021MNRAS.504.3375S} \textedit{because we anticipated that average dynamical lifetimes would increase due to the larger initial MOID with the inner planet.}  The dynamically colder orbit had a total of \textedit{12} unique simulations \textedit{defined using the same outer and inner planet masses used in the simulations of the hotter orbits}.\par
We then calculated the fractional error in energy \textedit{of the system}, the fractional error in the angular momentum \textedit{of the system}, and the difference between the initial and final mean longitudes for each planet using the same method used for the dynamically hotter orbits by integrating forward in time for 50 Myr and then backward in time for 50 Myr.

\section{Results}
\subsection{Dynamically Colder Orbit}
\textedit{Errors due to the integrator in planetary positions did not exceed order 10$^{-2}$ radian. The fractional error in energy and angular momentum did not exceed order 10$^{-10}$ for our dynamically colder orbit.}\par
 The normalized number of surviving test particles was negligible in every simulation and was zero for seven of the twelve simulations. The most likely fate of a test particle in a simulation was ejection, and the probability of a single test particle being ejected ranged from 67.83$\%$ to 83.49$\%$ over all simulations. In contrast, the maximum probability of a single test particle crossing the stellar Roche Limit was 0.83$\%$ over all dynamically cold simulations. The maximum probability of a test particle colliding with the inner planet was 15.39$\%$ and that of colliding with the outer planet was 17.33$\%$ for any one simulation. In almost every simulation, it was more likely for a test particle to collide with the inner planet than the outer planet. \par
\begin {figure} [htp]
\centering
\includegraphics[width = \columnwidth]{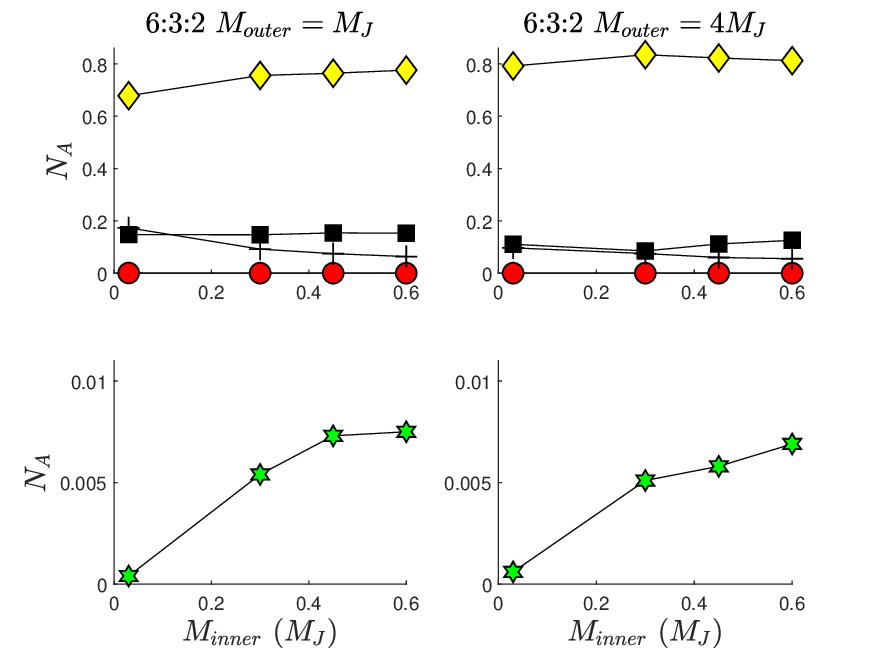}
\includegraphics[width = \columnwidth]{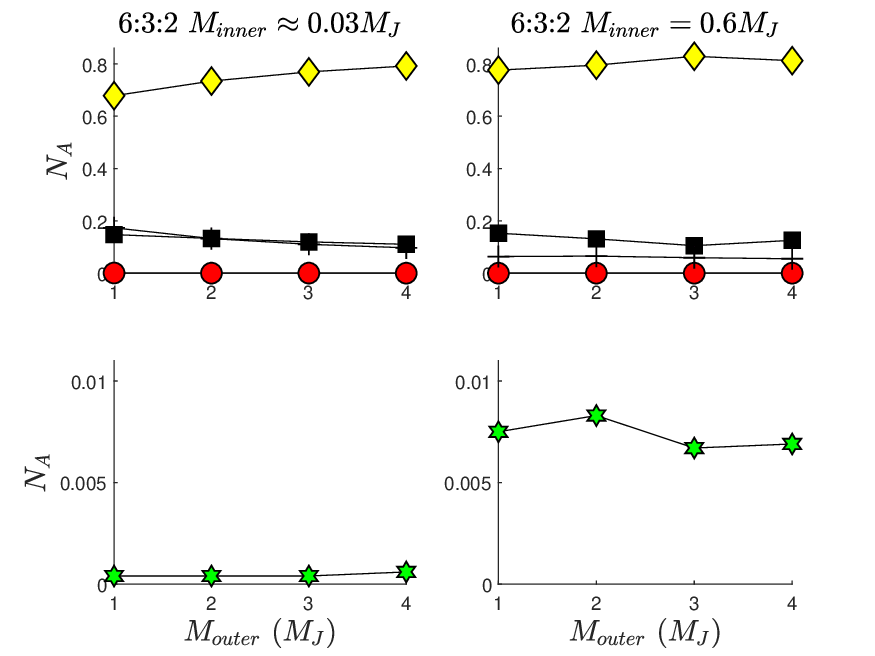}
\caption{Plots for the dynamically colder orbit. Top four panels: normalized number of test particles vs. inner planet mass for the 6:3:2 resonance for a constant outer planet mass for each test particle fate. The markers by fate are circle - survived the integration, diamond - ejection, star - crossed the stellar Roche Limit, square - collided with the inner planet, and plus sign - collided with the outer planet. The left panels are for a constant outer planet mass of $M_J$, and the right panels are for a constant outer planet mass of 4$M_J$. Bottom four panels: analogous plots of normalized number of test particles vs. outer planet mass with a constant inner planet mass of $\approx 0.03 M_J$ for the left panels and 0.6$M_J$ for the right panels.}
\label{fig_6_3_2_fates_plot_constant_outer_mass_100_Myr}
\end {figure}

 The top four panels in Figure \ref{fig_6_3_2_fates_plot_constant_outer_mass_100_Myr} show plots of the normalized number of test particles, $N_A$, vs. inner planet mass, $M_{inner}$, for a constant outer planet mass for each test particle fate. The left panels are for a constant outer planet mass of $M_J$, and the right panels are for a constant outer planet mass of $4M_J$.\par
\textedit{The markers used in the figure are by fate and are: circle - survived the integration, diamond - ejection, star - crossed the stellar Roche Limit, square - collided with the inner planet, and plus sign - collided with the outer planet. We will keep the same markers for each fate in corresponding plots for the dynamically hotter orbits discussed later.} \par
The top two panels show a very strong linear \textedit{anticorrelation} between the normalized number of test particles colliding with the outer planet and the inner planet mass. The linear regression coefficient is $< -0.95$ in each case. The correlation between the normalized number of test particles colliding with the inner planet and the inner planet mass was much weaker, and the linear regression coefficients of related plots in row 1 are 0.79 and 0.42 from left to right.\par 
The same was true for the correlation between the normalized number of test particles ejected and the inner planet mass. The linear regression coefficients of related plots in row 1 are 0.94 and 0.49 from left to right.\par
Of most interest in this work are the two plots in the second row which show a very strong linear correlation between the normalized number of test particles crossing the stellar Roche Limit and the inner planet mass. The linear regression coefficients are $> 0.95$ in each case. This is in stark contrast to the results of \citet[][]{2014MNRAS.439.2442F} who found that in single planet systems that planet mass is typically anticorrelated with the normalized number of test particles crossing the stellar Roche Limit. We determined that a typical accretion rate for a white dwarf star of 10$^8$ grams s$^{-1}$\citep[][]{2021ApJ...914...61K} could be explained by the accretion of an equivalent mass of one of our simulated asteroids every 13.8 Myr.\par

The bottom four panels show plots of the normalized number of test particles vs. outer planet mass, $M_{outer}$, for a constant inner planet mass of $\approx 0.03 M_J$ for the left panels and 0.6$M_J$ for the right panels for each test particle fate. There is a linear correlation between the normalized number of test particles ejected and the outer planet mass. The two related plots in row 3 have linear regression coefficients of  0.98 and 0.81 from left to right.\par
The normalized number of test particles colliding with the outer planet was strongly anticorrelated with the outer planet mass as the related plots in row 3 have linear regression coefficients of -0.97 and -0.89 from left to right which are both strong. There was somewhat of an anticorrelation between the normalized number of test particles colliding with the inner planet and the outer planet mass as the linear regression coefficients for the related plots in row 3 are -0.99 and -0.71 from left to right.\par
We can say with confidence that our data shows no consistent correlation between the normalized number of test particles crossing the stellar Roche Limit and the outer planet mass as the linear regression coefficients for the related plots in row 4 are 0.77 and -0.61 from left to right. \par
\begin{figure} [htp]
    \centering
    \includegraphics[width=\columnwidth]{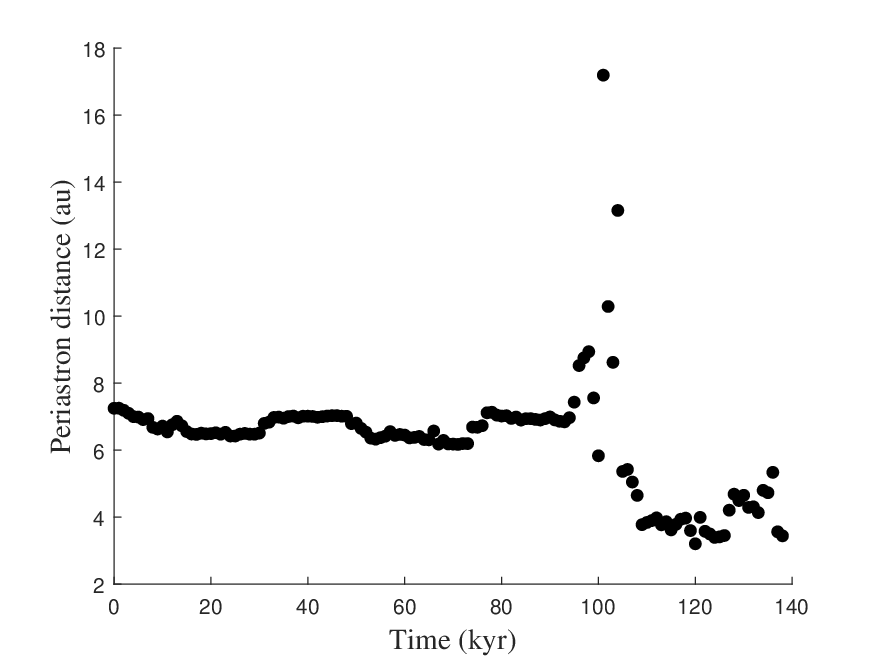}
    \includegraphics[width=\columnwidth]{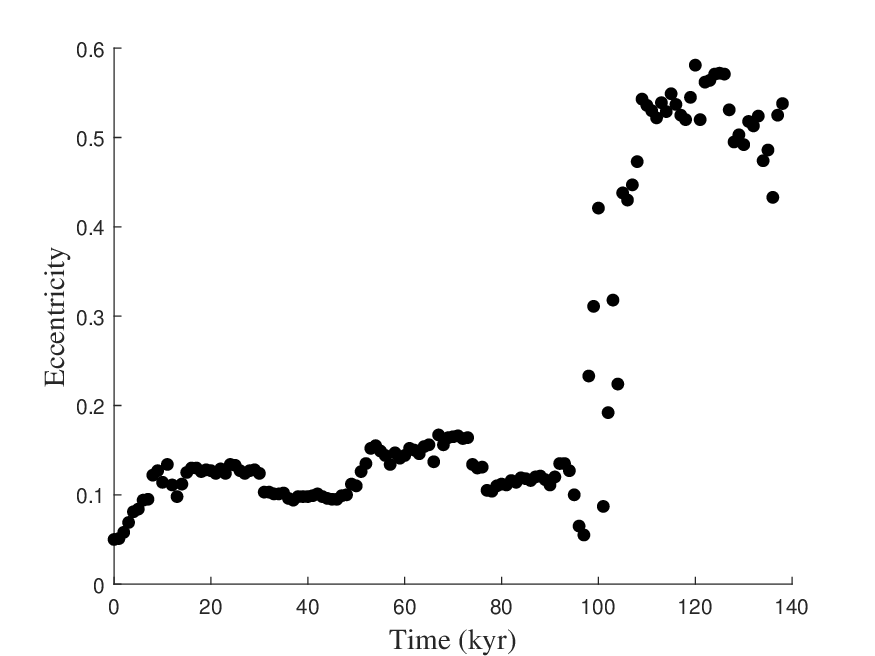}
    \caption{Top: A scatter plot of periastron distance vs. time for a test particle initially in the 6:3:2 resonance in the dynamically colder orbit that crossed the stellar Roche Limit. The outer planet mass was $M_J$, and the inner planet mass was $\approx0.03M_J$. Bottom: A scatter plot of eccentricity vs. time for the same test particle.}
    \label{fig_tp_5558_6_3_2_peri_v_time_100_Myr}
\end{figure}

For test particles that crossed the stellar Roche Limit, we witnessed eccentricity spikes that corresponded to times of periastron distance decreasing. Figure \ref{fig_tp_5558_6_3_2_peri_v_time_100_Myr} shows two scatter plots - one of periastron distance vs. time and the other of eccentricity vs. time for a test particle that crossed the stellar Roche Limit. Just after a time of 101 kyr, the periastron distance decreases dramatically from $\approx17$ au to below 4 au while the eccentricity spikes above 0.5. We found no consistent relationship between the normalized number of test particles and inner planet mass for the other test particle fates. 

\textedit{Average dynamical lifetimes for each simulation ranged from 23.0 kyr - 1137 kyr with a mean of 191 kyr over all 12 simulations.
Furthermore, we found strong exponential anticorrelations between the average dynamical lifetime of test particles in a simulation and both the inner and the outer planet mass. Overall, linear regression coefficients (that we call $R_{exp}$) for plots of the natural log of the average dynamical lifetime of test particles in a simulation vs. inner planet mass or outer planet mass ranged from -0.99 to -0.90.}\par
\begin{figure} [htp]
    \centering
    \includegraphics[width=\columnwidth]{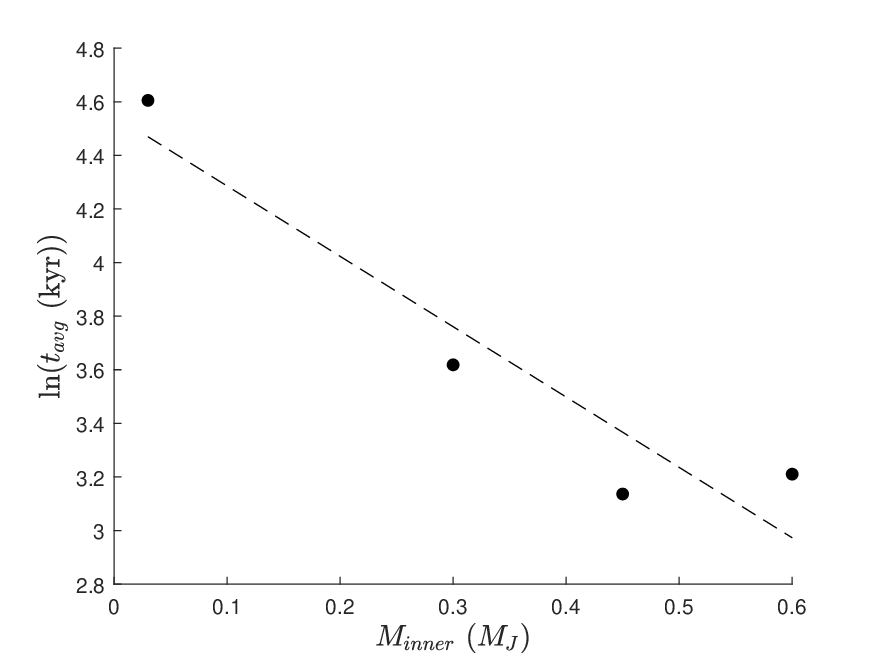}
    \includegraphics[width=\columnwidth]{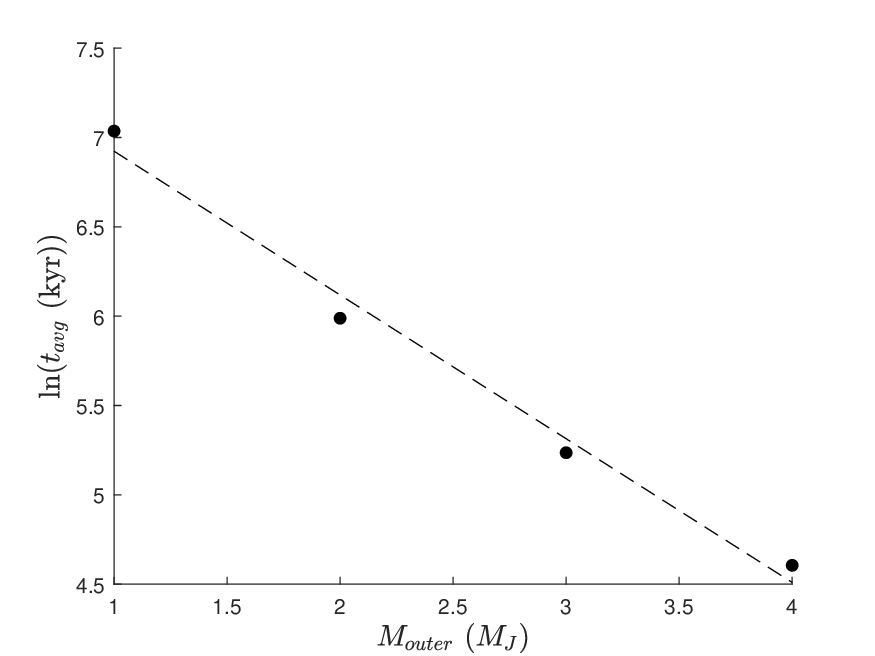}    
    \caption{\textedit{Top: the natural log of the average dynamical lifetime of test particles in a simulation vs. inner planet mass for a constant outer planet mass of $4M_J$ for the dynamically colder orbit. Bottom:  natural log of the average dynamical lifetime of test particles in a simulation vs. outer planet mass for a constant inner planet mass of $\approx 0.03M_J$. For both panels, the dashed line is the best-fit line.}}
\label{ln_avg_life_100_Myr}
\end{figure}

\textedit{As an example, Figure \ref{ln_avg_life_100_Myr} shows two plots: the top panel is a plot of the natural log of the average dynamical lifetime of test particles in a simulation vs. inner planet mass for a constant outer planet mass of $4M_J$, and the bottom panel is a plot of the natural log of the average dynamical lifetime of test particles in a simulation vs. outer planet mass for a constant inner planet mass of $\approx 0.03M_J$. For both panels, the dashed line is the best-fit line. The top plot has an $R_{exp}$ value of -0.94, and the bottom plot has an $R_{exp}$ value of -0.99.} 

Average dynamical lifetimes, $R_{exp}$ values, and the number of test particles that met each fate for each dynamically colder orbit simulation are shown in Table \ref{table_values_cold_6_3_2_Mi_6MJ} in the Appendix. 

\subsection{Dynamically Hotter Orbits}
\textedit{Errors  due to the integrator in planetary positions did not exceed order 10$^{-5}$ radian, and the fractional error in energy and angular momentum did not exceed order 10$^{-11}$ for dynamically hotter orbits.}\par
The normalized number of surviving test particles was negligible in every dynamically hotter simulation and was zero for eleven of the twenty four simulations. Table \ref{fate_probability_hotter} shows the range for the probability of a single test particle being ejected, crossing the stellar Roche Limit, or colliding with either planet for each resonance per simulation. There are some differences in the results between the two resonances, but overall we consider the results between the two resonances to be comparable.\par
The most likely fate of any test particle in a simulation was ejection. The probability of a single test particle being ejected for any one simulation ranged from 69.37$\%$ to 85.59$\%$. The odds of a single test particle colliding with a planet were below 19$\%$ for every simulation. Usually, it was more likely for a test particle to collide with the inner planet than the outer planet in a simulation. The maximum probability of a single test particle being delivered to the Roche Limit in a simulation was 1.08$\%$ for the 6:3:2 resonance and 0.91$\%$ for the 4:2:1 resonance. All of these probabilities are comparable to the those found for \textedit{the dynamically colder orbit.} \par
\begin{deluxetable} {cccccc}
\tablewidth{\columnwidth}
\tablecaption{Probability Ranges\label{fate_probability_hotter}}
\tablehead{    Res&Hi/Lo&Ejection&Roche&In. Pl&Out. Pl
}
\startdata    
4:2:1&Low&$69.82\%$&$0.05\%$&$8.98\%$&$3.94\%$\\  
4:2:1&High&$85.59\%$&$1.08\%$&$18.79\%$&$11.28\%$\\
6:3:2&Low&$69.37\%$&$0.03\%$&$8.09\%$&$5.16\%$\\
6:3:2&High&85.35\%&0.91\%&14.75\%&15.73\%\\
    \enddata
\tablecomments{\textedit{The range of the probability (from low to high) of a single test particle being ejected, crossing the stellar Roche Limit, or colliding with either planet for each resonance per simulation for dynamically hotter orbits. The columns are from left to right Res - the resonance, Hi/Lo - either the lower or upper bound of the range, Ejection - the percentage of test particles ejected from the system, Roche - the percentage of test particles that crossed the stellar Roche Limit, In. Pl - the percentage of test particles that collided with the inner planet, Out. Pl - the percentage of test particles that collided with the outer planet.}}
\end{deluxetable}

The 4:2:1 resonance sent 1.1 times more test particles to the stellar Roche Limit than the 6:3:2 resonance. The 4:2:1 resonance sent 734 test particles to the stellar Roche Limit over its twelve simulations while the 6:3:2 resonance sent only 649. \par
The hotter 6:3:2 resonance sent 1.2 times as many test particles to the stellar Roche Limit than the colder 6:3:2 resonance over twelve simulations. In each simulation to analogous simulation comparison (hot compared to cold), the hotter orbit delivered more test particles to the stellar Roche Limit in eleven out of twelve cases. In the one exception, the colder orbit delivered four test particles to the Roche Limit while the hotter orbit only delivered three. Thus, an asteroid in a more eccentric orbit generally has a higher probability of being delivered to the stellar Roche Limit than an asteroid in a less eccentric orbit all other factors being the same. \par
\begin {figure} [htp]
\centering
\includegraphics[width = \columnwidth]{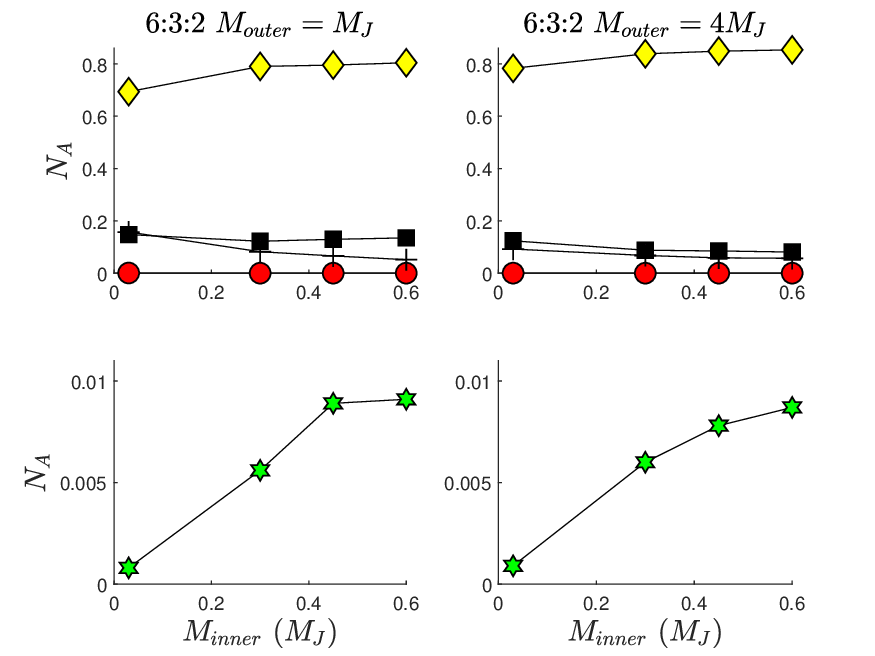}
\includegraphics[width = \columnwidth]{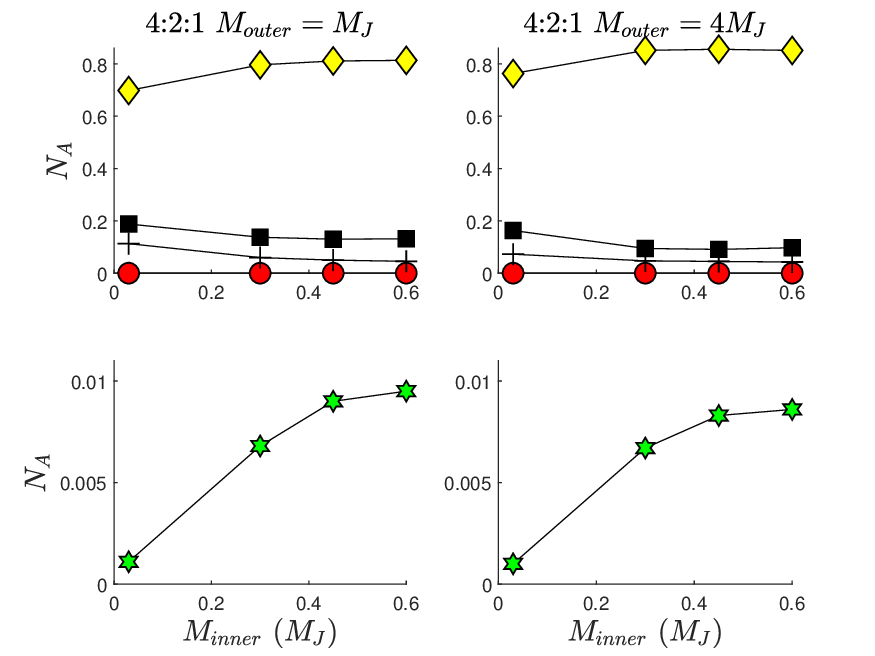}
\caption{Plots for dynamically hotter orbits. Top four panels: Normalized number of test particles vs. inner planet mass for the 6:3:2 resonance for a constant outer planet mass for each test particle fate. The markers by fate are circle - survived the integration, diamond - ejected, star - crossed the stellar Roche Limit, square - collided with the inner planet, and plus sign - collided with the outer planet. The left panels are for a constant outer planet mass of $M_J$, and the right panels are for a constant outer planet mass of 4$M_J$. Bottom four panels: analogous plots for the 4:2:1 resonance.}
\label{fig_6_3_2_fates_plot_constant_outer_mass_10_Myr}
\end {figure}

The top four panels in Figure \ref{fig_6_3_2_fates_plot_constant_outer_mass_10_Myr} show plots of the normalized number of test particles vs. inner planet mass for a constant outer planet mass for each test particle fate. The left panels are for a constant outer planet mass of $M_J$, and the right panels are for a constant outer planet mass of $4M_J$ for dynamically hotter orbits for both resonances. As with the dynamically colder orbit, the normalized number of test particles delivered to the stellar Roche Limit was strongly correlated with the inner planet mass for both resonances with linear regression coefficients all above 0.9 for each of the plots in rows 2 and 4. \par
The normalized number of test particles colliding with the inner planet was anticorrelated with the inner planet mass for the 4:2:1 resonance as linear regression coefficients of related plots are -0.90 and -0.84 from left to right in row 1, but for the 6:3:2 resonance the anticorrelation is strong only when the outer planet mass is $4M_J$ as linear regression coefficients are -0.51 and -0.93 from left to right in row 3.\par
The normalized number of test particles colliding with the outer planet was strongly anticorrelated with inner planet mass as we also saw with \textedit{the dynamically colder orbit}. The linear regression coefficients for related plots in rows 1 and 3 are all below -0.91. The normalized number of test particles ejected was linearly correlated with the inner planet mass as demonstrated by the linear regression coefficients of 0.92 and 0.86 for the 4:2:1 resonance; and 0.91 and 0.94 for the 6:3:2 resonance from left to right for related plots in rows 1 and 3, respectively.\par
\begin {figure} [H]
\centering
\includegraphics[width = \columnwidth]{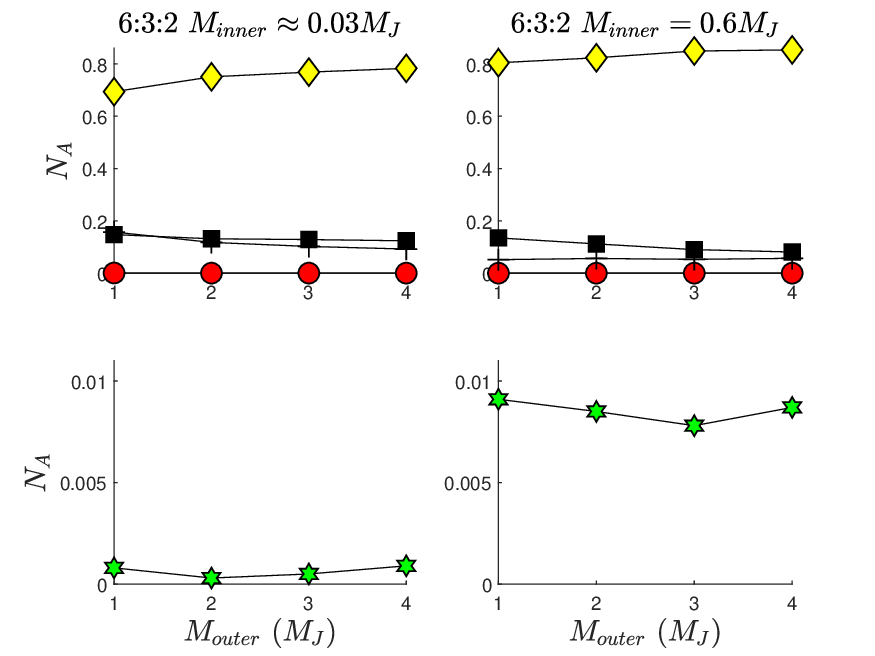}
\includegraphics[width = \columnwidth]{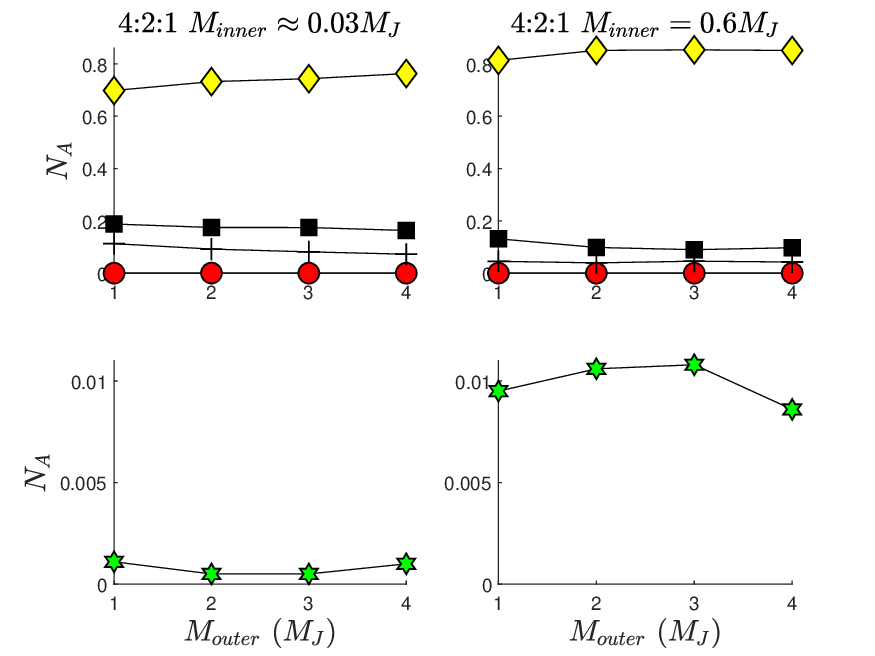}
\caption{Plots for dynamically hotter orbits. Top four panels: Normalized number of test particles vs. outer planet mass for the 6:3:2 resonance for a constant inner planet mass for each test particle fate. The markers by fate are circle - survived the integration, diamond - ejected, star - crossed the stellar Roche Limit, square - collided with the inner planet, and plus sign - collided with the outer planet. The left panels are for a constant inner planet mass of $\approx0.03M_J$, and the right panels are for a constant inner planet mass of 0.6$M_J$. Bottom four panels: analogous plots for the 4:2:1 resonance.}
\label{fig_6_3_2_fates_plot_constant_inner_mass_10_Myr}
\end {figure}

Figure \ref{fig_6_3_2_fates_plot_constant_inner_mass_10_Myr} shows plots of the normalized number of test particles vs. outer planet mass for a constant inner planet mass of $\approx 0.03 M_J$ for the left panels and 0.6$M_J$ for the right panels for each test particle fate for dynamically hotter orbits for both of our selected resonances. The linear correlation between the normalized number of test particles ejected and the outer planet mass was mostly very strong. The related plots in row 1 for the 6:3:2 resonance have linear regression coefficients above 0.94 in both cases which is strong. The related plots for the 4:2:1 resonance in row 3 have linear regression coefficients of 0.98 and 0.77 from left to right. The four plots in rows 2 and 4 show that there is no correlation between the normalized number of test particles that crossed the stellar Roche Limit and outer planet mass for either resonance. Linear regression coefficients over these four plots ranged from -0.45 to 0.23 which are weak and contradictory correlations.\par
The normalized number of test particles colliding with the inner planet was anticorrelated with the outer planet mass for both resonances. For the 6:3:2 resonance, the related plots in row 1 have linear regression coefficients of -0.93 and -0.98 from left to right. For the 4:2:1 resonance, the related plots in row 3 have linear regression coefficients of -0.94 and -0.77 from left to right.
\par
The normalized number of test particles colliding with the outer planet sometimes showed a strong anticorrelation with the outer planet mass. The related plots when the inner planet mass was $\approx 0.03 M_J$ in rows 1 and 3 have linear regression coefficients of -0.95 and -0.98, respectively, but the related plots when the inner planet mass was $0.6M_J$ have linear regression coefficients of 0.61 and -0.06, respectively.

\begin{figure}
    \centering
    \includegraphics[width=\columnwidth]{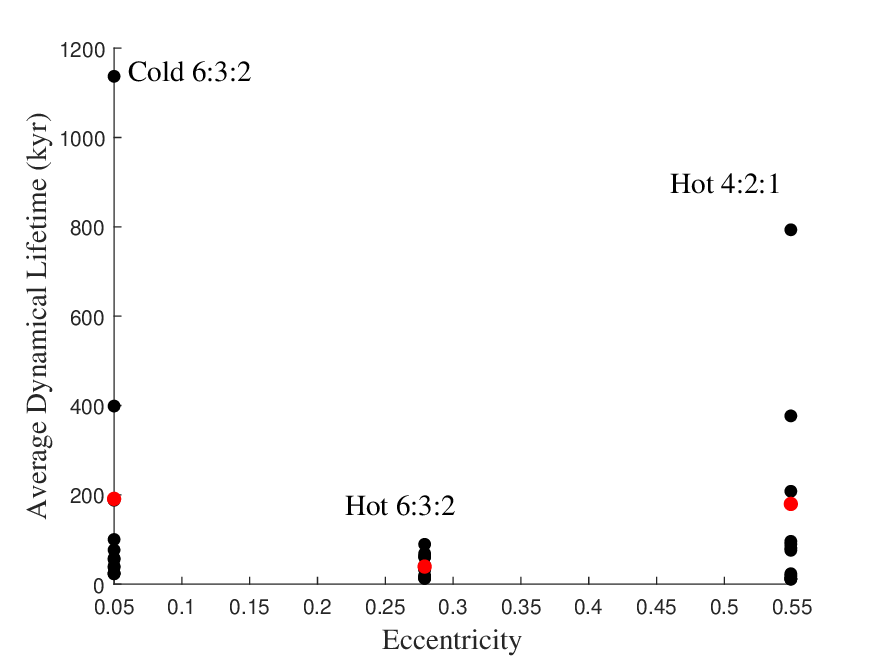}
    \caption{\textedit{Average dynamical lifetime vs. eccentricity for the dynamically colder (far left) and hotter orbits for all 12 simulations for each orbit. Times for the hotter orbit in the 6:3:2 resonance are in the midlde while those for the orbit in the hotter 4:2:1 resonance are on the right. Mean average dynamical lifetimes for each orbit are plotted as red circles.}}
    \label{avg_dynamical_lifetimes}
\end{figure}

\textedit{Average dynamical lifetimes ranged from 10.8 kyr - 793.4 kyr for the orbit in 4:2:1 resonance and from 12.9 kyr - 89.2 ky for the orbit in the 6:3:2 resonance. Over all 12 simulations for each orbit, the mean average dynamical lifetime was 179.5 kyr for the orbit in 4:2:1 resonance and 39.3 kyr for the orbit in 6:3:2 resonance. This difference is due to the orbit in the 4:2:1 resonance having a greater initial eccentricity and lower initial semi-major axis than those of the orbit in the 6:3:2 resonance. Because of this, the initial periastron distance of the orbit in the 4:2:1 resonance was 2.66 au less than the initial periastron distance of the orbit in the 6:3:2 resonance. Thus, test particles in the 4:2:1 resonance could be at farther distances from the inner planet compared to those in the 6:3:2 resonance and feel less perturbation. The mean average dynamical lifetime over 12 simulations for the colder orbit is an order of magnitude larger than that of the hotter orbit in the 6:3:2 resonance. The reason for this is the smaller initial MOID of the colder orbit. The dynamical lifetimes for each hotter orbit and the colder orbit are plotted against eccentricity in Figure \ref{avg_dynamical_lifetimes}. Mean average dynamical lifetimes for each of the 12 simulations for each orbit are plotted as red circles. }\par

As with the dynamically colder \textedit{orbit, average dynamical lifetimes were usually strongly exponentially anticorrelated with both the inner and the outer planet mass. For anticorrelations, $R_{exp}$ values found using either the inner or outer planet mass planet mass ranged from $\approx$-0.99 to -0.85 over both resonances.}\par 
\textedit{However, there was one case for which there was a strong positive correlation. It was the case involving simulations for the 4:2:1 resonance for which the outer planet mass was a constant $M_J$. In this one case, the $R_{exp}$ value was +0.99.}\par
\textedit{This was due to competing factors. As the inner planet mass increased, gravitational perturbations of the inner planet on the test particles increased, thus decreasing dynamical lifetimes, however, an increased mass also reduced the perturbing effect of the outer planet mass on the inner planet mass thus decreasing the eccentricity range of the inner planet which would increase dynamical lifetimes.}\par
\textedit{Average dynamical lifetimes, $R_{exp}$ values, and the number of test particles that met each fate for each dynamically hotter orbit simulation are shown in Tables \ref{table_values_hot_6_3_2} and \ref{table_values_hot_4_2_1} in the Appendix.}\par

\newpage
\section{Conclusions}
In this work, we used numerical integration of the 4-body problem to study \textedit{3-body resonance chains} (two planets and an asteroid in the innermost orbit) as a possible mechanism for white dwarf pollution.\par
Two \textedit{3-body resonance chains} were selected for study: the 6:3:2 resonance and the 4:2:1 resonance. Asteroids in a dynamically colder initial orbit in the 6:3:2 resonance and hotter initial orbits in both resonances were studied. In the dynamically colder orbit, an asteroid had up to a 0.83$\%$ chance of being delivered to the stellar Roche Limit. This probability was strongly linearly correlated with the mass of the inner planet but was not correlated with the mass of the outer planet. The most likely fate of an asteroid in the dynamically colder orbit was ejection from the planetary system. \textedit{Average dynamical lifetimes ranged from 23 kyr to 1137 kyr for the dynamically colder orbit, and the average dynamical lifetime was exponentially anticorrelated with both the inner and the outer planet mass.}\par
For dynamically hotter orbits, the maximum probability of an asteroid being delivered to the stellar Roche Limit was 1.08$\%$ for the 4:2:1 resonance and was 0.91$\%$ for the 6:3:2 resonance. The hotter 6:3:2 resonance delivered 1.2 times more asteroids to the stellar Roche Limit than the colder 6:3:2 resonance. This implies that an asteroid in a more eccentric orbit has a higher probability of being delivered to the stellar Roche Limit than an asteroid in a less eccentric orbit, all other factors being the same.\par
\textedit{For dynamically hotter orbits,} the 4:2:1 resonance delivered 1.1 times more asteroids to the stellar Roche Limit than the 6:3:2 resonance. The probability of an asteroid being delivered to the stellar Roche Limit from a dynamically hotter orbit was strongly linearly correlated with the mass of the inner planet for both resonances but was not correlated with the mass of the outer planet.\par
\textedit{For dynamically hotter orbits,} the most likely fate of an asteroid was ejection from the planetary system. \textedit{Average dynamical lifetimes ranged from 12.9 kyr to 89.2 kyr for the 6:3:2 resonance and from 10.8 kyr to 793.4 kyr for the 4:2:1 resonance.}\par
 \textedit{The mean average dynamical lifetime was 179.5 kyr for the 4:2:1 resonance and 39.3 kyr for the 6:3:2 resonance. This difference is due to the greater eccentricity and lower semi-major axis of the test particles' initial orbit for the 4:2:1 resonance which lowered the periastron distance, allowing test particles to be at farther distances from the inner planet and feel less perturbation compared to test particles in the 6:3:2 resonance.}\par
\textedit{Average dynamical lifetimes were usually strongly exponentially anticorrelated with both the inner and the outer planet mass. However, there was one case for which there was a strong exponential positive correlation with inner planet mass. This was due to competing factors. As the inner planet mass increased, gravitational perturbations due to the inner planet on the asteroids increased, decreasing dynamical lifetimes, however, an increased mass also reduced the perturbing effect of the outer planet mass on the inner planet mass, increasing dynamical lifetimes due to the reduced inner planet eccentricity range.}\par 
A typical accretion rate for a white dwarf star of 10$^8$ grams s$^{-1}$ could be explained by the accretion of an equivalent mass of one of our simulated asteroids every 13.8 Myr. In future work, we will study the efficiency of different resonances on delivering an asteroid to the stellar Roche Limit.

\newpage
\setcounter{secnumdepth}{0} 
\section{Acknowledgements}
This project made use of the Jetstream 2 computer at Indiana University. The authors would like to thank the referee for their comments which greatly improved this work.

\clearpage
\appendix


\begin{deluxetable*}{cccccccccc}[ht]
    %
    \tablecaption{\\Colder Orbit Simulations 6:3:2 resonance\label{table_values_cold_6_3_2_Mi_6MJ}}
\tablehead{\colhead{$M_{inner}$}&\colhead{$M_{outer}$}&\colhead{$t_{avg}$}&\colhead{$R_{exp}$}&\colhead{Survived}&\colhead{Ejection}&\colhead{Roche}&\colhead{Inner Planet}&\colhead{Outer Planet}\\
\colhead{($M_J$)}&\colhead{($M_J$)}&\colhead{(kyr)}&&&&&}

\startdata
$\approx 0.03M_J$&1&1137.0&-0.90&7&6783&4&1473&1733\\
0.3&1&76.6&...&1&7558&54&1468&919\\
0.45&1&55.2&...&0&7642&73&1539&746\\
0.6&1&57.9&...&0&7761&75&1528&636\\
$\approx 0.03M_J$&4&100.0&-0.94&0&7920&6&1105&969\\
0.3&4&37.3&...&0&8349&51&850&750\\
0.45&4&23.0&...&0&8225&58&1117&600\\
0.6&4&24.8&...&0&8123&69&1256&552\\
$\approx 0.03M_J$&1&1137.0&-0.99&7&6783&4&1473&1733\\
$\approx 0.03M_J$&2&398.7&...&1&7343&4&1329&1323\\
$\approx 0.03M_J$&3&187.8&...&3&7694&4&1195&1104\\
$\approx 0.03M_J$&4&100.0&...&0&7920&6&1105&969\\
0.6&1&57.9&-0.93&0&7761&75&1528&636\\
0.6&2&40.1&...&1&7952&83&1312&652\\
0.6&3&23.6&...&0&8290&67&1052&591\\
0.6&4&24.8&...&0&8123&69&1256&552\\
\enddata

    \begin{flushleft}
        
    \tablecomments{Tables of values for simulations of the dynamically colder orbit in the 6:3:2 resonance over 100 Myr. $M_{inner}$ and $M_{outer}$ are the masses of the inner and outer planet in units of Jupiter masses, respectively. $t_{avg}$ is the average dynamical lifetime. $R_{exp}$ is the linear regression coefficient for a plot of the natural log of the average lifetime vs either the inner planet mass or outer planet mass. Rows are shown in groups of four in which either the inner or the outer planet mass is held constant. The values of the planet's mass that change within a group are used to find the $R_{exp}$ value in the first row that starts a group of four rows. The final five columns are for the number of test particles that met each fate. From left to right these columns are Survived - the number that survived the integration, Ejection - the number ejected from the system, Roche - the number that crossed their stellar Roche Limit, Inner Planet - the number that collided with the inner planet, and Outer Planet - the number that collided with the outer planet. Some rows are repeated for convenience.}
        \end{flushleft}

\end{deluxetable*}


\begin{deluxetable*}{cccccccccc}[ht]
    \tablewidth{0pt}
    
    \tablecaption{\\Hotter Orbit Simulations 6:3:2 resonance\label{table_values_hot_6_3_2}}
\tablehead{\colhead{$M_{inner}$}&\colhead{$M_{outer}$}&\colhead{$t_{avg}$}&\colhead{$R_{exp}$}&\colhead{Survived}&\colhead{Ejection}&\colhead{Roche}&\colhead{Inner Planet}&\colhead{Outer Planet}\\
\colhead{($M_J$)}&\colhead{($M_J$)}&\colhead{(kyr)}&&&&&}

\startdata
$\approx 0.03M_J$&1&89.2&-0.85&7&6937&8&1475&1573\\
0.3&1&33.7&...&0&7903&56&1221&820\\
0.45&1&42.0&...&1&7955&89&1295&660\\
0.6&1&33.1&...&1&8043&91&1349&516\\
$\approx 0.03M_J$&4&60.9&-0.92&2&7830&9&1237&922\\
0.3&4&16.2&...&0&8385&60&878&677\\
0.45&4&13.3&...&1&8487&78&849&585\\
0.6&4&12.9&...&1&8535&87&809&568\\
$\approx 0.03M_J$&1&89.2&-0.91&7&6937&8&1475&1573\\
$\approx 0.03M_J$&2&68.4&...&2&7505&3&1314&1176\\
$\approx 0.03M_J$&3&61.3&...&3&7685&5&1285&1022\\
$\approx 0.03M_J$&4&60.9&...&2&7830&9&1237&922\\
0.6&1&33.1&-0.95&1&8043&91&1349&516\\
0.6&2&19.6&...&0&8233&85&1119&563\\
0.6&3&14.1&...&1&8492&78&901&528\\
0.6&4&12.9&...&1&8535&87&809&568\\
\enddata
    \begin{flushleft}

    \tablecomments{Tables of values for simulations of the dynamically hotter orbit in the 6:3:2 resonance over 10 Myr. $M_{inner}$ and $M_{outer}$ are the masses of the inner and outer planet in units of Jupiter masses, respectively. $t_{avg}$ is the average dynamical lifetime. $R_{exp}$ is the linear regression coefficient for a plot of the natural log of the average lifetime vs either the inner planet mass or outer planet mass. Rows are shown in groups of four in which either the inner or the outer planet mass is held constant. The values of the planet's mass that change within a group are used to find the $R_{exp}$ value in the first row that starts a group of four rows. The final five columns are for the number of test particles that met each fate. From left to right these columns are Survived - the number that survived the integration, Ejection - the number ejected from the system, Roche - the number that crossed their stellar Roche Limit, Inner Planet - the number that collided with the inner planet, and Outer Planet - the number that collided with the outer planet. Some rows are repeated for convenience.}
        \end{flushleft}

\end{deluxetable*}
\begin{deluxetable*}{cccccccccc}[ht]
    \tablewidth{0pt}

    \tablecaption{\\Hotter Orbit Simulations 4:2:1 resonance\label{table_values_hot_4_2_1}}
\tablehead{\colhead{$M_{inner}$}&\colhead{$M_{outer}$}&\colhead{$t_{avg}$}&\colhead{$R_{exp}$}&\colhead{Survived}&\colhead{Ejection}&\colhead{Roche}&\colhead{Inner Planet}&\colhead{Outer Planet}\\
\colhead{($M_J$)}&\colhead{($M_J$)}&\colhead{(kyr)}&&&&&}

\startdata
$\approx 0.03M_J$&1&95.8&0.99&0&6982&11&1879&1128\\
0.3&1&207.9&...&0&7963&68&1373&596\\
0.45&1&376.7&...&0&8112&90&1300&498\\
0.6&1&793.4&...&2&8135&95&1314&454\\
$\approx 0.03M_J$&4&75.7&-0.96&1&7631&10&1632&726\\
0.3&4&18.5&...&1&8517&67&945&470\\
0.45&4&11.8&...&0&8559&83&909&449\\
0.6&4&10.8&...&0&8513&86&973&428\\
$\approx 0.03M_J$&1&95.8&-0.99&0&6982&11&1879&1128\\
$\approx 0.03M_J$&2&89.2&...&5&7322&5&1747&921\\
$\approx 0.03M_J$&3&81.0&...&2&7433&5&1745&815\\
$\approx 0.03M_J$&4&75.7&...&1&7631&10&1632&726\\
0.6&1&793.4&-0.86&2&8135&95&1314&454\\
0.6&2&23.3&...&0&8516&106&984&394\\
0.6&3&12.5&...&0&8536&108&898&458\\
0.6&4&10.8&...&0&8513&86&973&428\\
\enddata
    \begin{flushleft}

    \tablecomments{Tables of values for simulations of the dynamically hotter orbit in the 4:2:1 resonance over 10 Myr. $M_{inner}$ and $M_{outer}$ are the masses of the inner and outer planet in units of Jupiter masses, respectively. $t_{avg}$ is the average dynamical lifetime. $R_{exp}$ is the linear regression coefficient for a plot of the natural log of the average lifetime vs either the inner planet mass or outer planet mass. Rows are shown in groups of four in which either the inner or the outer planet mass is held constant. The values of the planet's mass that change within a group are used to find the $R_{exp}$ value in the first row that starts a group of four rows. The final five columns are for the number of test particles that met each fate. From left to right these columns are Survived - the number that survived the integration, Ejection - the number ejected from the system, Roche - the number that crossed their stellar Roche Limit, Inner Planet - the number that collided with the inner planet, and Outer Planet - the number that collided with the outer planet. Some rows are repeated for convenience.}
        \end{flushleft}

\end{deluxetable*}


\end{document}